\documentclass[11pt,letterpaper]{article}
\usepackage[T1]{fontenc}
\usepackage[utf8]{inputenc}
\usepackage{lmodern,microtype,amsmath,amssymb,graphicx,booktabs,tabularx,array}
\usepackage[left=.94in,right=.94in,top=.83in,bottom=.82in,headheight=14pt,headsep=12pt]{geometry}
\usepackage{float,placeins,caption,fancyhdr}
\usepackage[numbers,sort&compress]{natbib}
\usepackage[hidelinks]{hyperref}
\hypersetup{pdftitle={Forensic Paired Auditing of Multi-Hop Retrieval with Pair-ID},pdfauthor={Wenzhang Du},pdfsubject={Paired evidence interventions and retrieval decisions}}

\newcolumntype{L}{>{\raggedright\arraybackslash}X}
\newcommand{\pname}[1]{\textsf{\footnotesize #1}}

\title{Forensic Paired Auditing of Multi-Hop Retrieval with Pair-ID}
\author{Wenzhang Du}
\date{}
\makeatletter
\renewcommand{\maketitle}{%
  \begin{center}
  \begin{minipage}{.89\linewidth}\centering\LARGE\bfseries\@title\par\end{minipage}\par
  \vspace{8pt}{\large\@author\par}
  {\small Independent Researcher, China\par}
  \end{center}}
\makeatother
\begin{document}
\thispagestyle{plain}
\maketitle
\begin{abstract}
A multi-hop retrieval system can acquire a correct answer and still return an incorrect one. We introduce Pair-ID, a paired audit that tracks how evidence edits, retrieval actions, and answer return contribute to the final score. We cross support addition with passage removal, then test retrieval feedback on a separate sample by restoring the original context and fixing five candidate actions. On 1,024 held-out HotpotQA and 2WikiMultiHopQA questions, temporary deletion and intact-context diagnosis yielded different actions in 45.4\% of question--seed pairs. With direct return, their exact-match difference was $+0.33$ percentage points (95\% interval $-0.55$ to $+1.21$). The targeted selector with a verifier trailed fixed challenge search by 8.72 points and used 0.261 more allocated GPU seconds per question. In a post-hoc ablation on the same selected actions, the verifier lowered exact match by 8.11 points. With direct return on both paths, the targeted selector was 0.62 points below fixed search.
\end{abstract}

\section{Introduction}
\label{sec:intro}
A retrieval agent can edit the evidence, choose a new search, and override the answer produced by that search. A failed search and an override that replaces a correct answer with a wrong one both earn zero exact match. \textbf{Pair-ID} records the answers available under each evidence edit and retrieval action, then measures which answer the system actually returns.

\noindent We make three contributions.
\begin{list}{\textbullet}{\setlength{\leftmargin}{1.5em}\setlength{\topsep}{3pt}\setlength{\itemsep}{2pt}\setlength{\parsep}{0pt}\setlength{\partopsep}{0pt}}
\item \textbf{A paired evidence-edit test.} A $2\times2$ design identifies 89 joint-only repairs. Length-matched controls and two readers test how the result depends on passage content and scoring (Section~\ref{sec:edits}).
\item \textbf{A controlled feedback test.} Given the restored passages, both selectors choose among five actions before seeing their answers. We measure how often a different choice changes the score (Section~\ref{sec:decisions}).
\item \textbf{An answer-return ablation.} On the targeted selector's own choices, allowing the verifier to override the terminal answer costs 8.11 exact-match points (Section~\ref{sec:return}).
\end{list}

\begin{figure}[H]
\centering
\includegraphics[width=\linewidth]{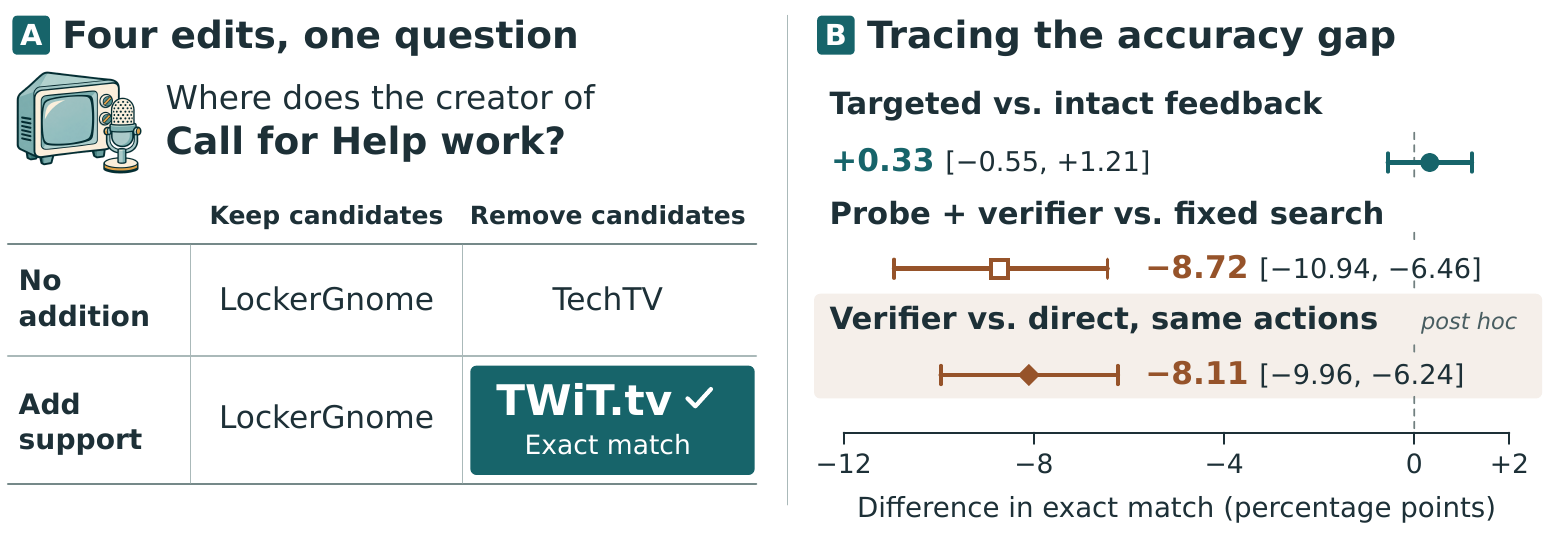}
\caption{(A) Only the joint edit repairs this question, the first such case in sample-hash order. (B) Three comparisons on a separate 1,024-question test. The top row varies feedback with direct return. The middle row changes policy and return rule. The bottom row holds the targeted actions fixed and changes only the return rule (post hoc). Bars show pointwise 95\% group-bootstrap intervals.}
\label{fig:first}
\end{figure}

What can an edit reveal? On a 2WikiMultiHopQA question about the creator of \emph{Call for Help}, both the unedited reader and the reader given missing support named LockerGnome. Deleting candidate passages produced TechTV. Only the joint edit yielded TWiT.tv (Figure~\ref{fig:first}A). The labels that identify this successful edit are available to the evaluator. A system choosing its next action must work from the question, its retrieved passages, and its own answer.

Existing diagnostics measure the quality of retrieved evidence and generated answers \citep{ru2024ragchecker,jain2026cuer}, and adaptive systems learn when to change retrieval strategy \citep{jeong2024adaptive}. Pair-ID connects these tasks by holding each candidate action's evidence and answer fixed when comparing two feedback conditions. It then holds the chosen action fixed to test what the verifier returns. We evaluate these comparisons on HotpotQA and 2WikiMultiHopQA, whose support annotations also permit a direct check of which evidence edits helped each reader \citep{yang2018hotpotqa,ho2020constructing}.

\section{Pair-ID audit design}
\label{sec:methods}
\paragraph{Measuring the response to an evidence edit.}
Let $S$ be the ordered initial passages for a question, $G$ a set to append and $D\subseteq S$ a set of removal candidates. Let $Y_{ad}$ be the exact-match score of a reader given $S$, $S+G$, $S\setminus D$ or $(S\setminus D)+G$, for $(a,d)=(0,0),(1,0),(0,1),(1,1)$ respectively. Surviving passages retain their order. The vector $V=(Y_{10},Y_{01},Y_{11})$ records what happened under each edit. For an initially incorrect answer, $V=(0,0,1)$ means that only the combination succeeded. For annotation-guided addition, $G$ is empty when all annotated support is present and the two addition cells repeat their corresponding prompts. Equal prompts share one generated answer within the same reader and decoding replicate.

\paragraph{Benchmark contexts and closed-corpus retrieval.}
In the benchmark-context experiment, BM25 ranked each question's supplied passages. Missing support \emph{sentences} were appended to its top five. In a second study BM25 searched the full collection of public training and development passages for each benchmark. The reader again saw the top five, and annotation-guided addition now appended missing support \emph{paragraphs}. Removal excluded annotated support titles, support sentences and answer strings or aliases. A natural language inference model screened the remaining candidates against the reference and annotated support. An entailment prediction or an overlength hypothesis pair protected the passage from removal. These filters can miss a bridge or an alternate reasoning chain.

A feasible editor uses only the question and retrieved text. Its addition appended the two passages with highest relevance scores among BM25 ranks 6--50. Feasible deletion pruned the two lowest-scoring passages in the starting five. A BGE reranker scored the question against overlapping windows covering each whole passage \citep{xiao2023cpack}. We ran both sets of edits on Qwen3-4B, with thinking disabled, and Mistral-7B-Instruct-v0.3 \citep{yang2025qwen3,jiang2023mistral}. Both used their own chat templates and a common short-answer instruction. We scored the greedy baseline first. Each reader's greedy baseline score defined its failure cohort for the repeated measurements.

\begin{figure}[htbp]
\centering
\includegraphics[width=\linewidth]{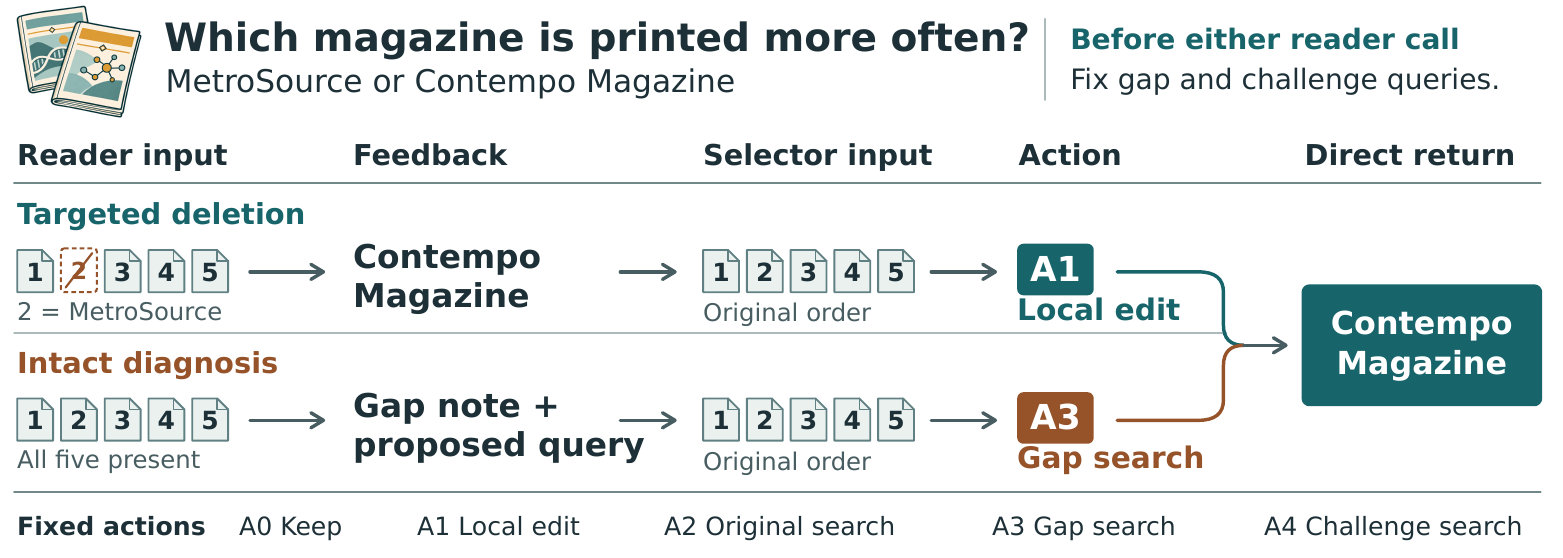}
\caption{Two choices on the HotpotQA test question with the smallest sample hash, using the first fitting seed and feedback draw. Passage 2, \emph{MetroSource}, is withheld only from the targeted reader call. Both selectors see the original five passages. One chooses a local edit ($A_1$), the other the precommitted gap search ($A_3$). The diagnosis's proposed query is an input to the selector. Both actions return the correct answer.}
\label{fig:reset}
\end{figure}

\paragraph{Temporary deletion before a retrieval decision.}
A separate Qwen experiment starts from the five BM25 passages. A planner first commits a query for missing evidence (a \emph{gap search}) and a query seeking evidence that could challenge the initial answer (a \emph{challenge search}). In the targeted condition, question and initial answer together rank the original passages. The highest-ranked passage is removed for one sampled reader call. The selector receives that response alongside the five original passages in their original order. In the principal comparison, an equally available reader call inspects the intact evidence and describes a possible gap. Its proposed search string can inform the selector but cannot alter the fixed action menu. The final reader sees the question and the chosen action's context. References and support labels enter training and offline scoring. Figure~\ref{fig:reset} follows both feedback paths on one question.

\paragraph{Five possible actions and two return rules.}
The selector can keep the original answer ($A_0$), prune and append locally ($A_1$), search more widely with the original question ($A_2$), or run the committed gap ($A_3$) or challenge query ($A_4$). Search combines BM25 and dense retrieval, reranks the candidates, retains up to three original passages and appends up to two new ones. For each question $q$ we record the answer $t_q(a)$ produced by each action $a$. A selector using feedback $f$ chooses $a_f$ before it can inspect these answers. Direct return scores $t_q(a_f)$ against the reference. For the return-rule ablation we hold $a_f$ and $t_q(a_f)$ fixed and let the verifier choose among the initial, probe, and terminal answers. A separate diagnostic executes the query the intact reader proposes. It can retrieve different passages from $A_3$, whose query was committed before feedback.

\paragraph{Evaluation protocol.}
The two retrieval studies reuse 1,024 training and 256 validation questions. Their tests comprise disjoint groups. Within each benchmark, questions sharing normalized question text or a complete unordered set of support titles are linked before partitioning. The acquisition test uses 512 questions per dataset, in 464 HotpotQA and 407 2Wiki groups. In the principal contrast, we pair the policies on the same question, initial state, feedback draw and fitting seed. We average the three seed-specific differences within question, then give each dataset mean equal weight. Pointwise 95\% intervals resample whole groups within each dataset 10,000 times, with equal weight per question inside each draw. These intervals describe question sampling for the fitted policies. Repair rates in benchmark contexts use the failures eligible for the corresponding edit. Table~\ref{tab:inventory} gives the overlap and role of each sample.

\section{What paired evidence edits reveal}
\label{sec:edits}
\paragraph{Some repairs require both edits.}
Among 1,190 initially incorrect answers to questions with benchmark-provided contexts, 89 became correct only when we both added missing support and removed candidate passages. Neither edit alone repaired these answers. In 769 cases, all three edited answers remained incorrect. The same initially wrong answer can therefore mask either a joint-only repair or failure under every tested edit.

Addition alone repaired 197 of the 600 failures with missing annotated support (32.8\%, 95\% interval 29.2--36.7). Deletion alone repaired 53 of those 600 and 109 of the 590 failures whose context already contained all annotated support. Across all 1,190 failures, its repair rate was 13.6\% (162/1,190, interval 11.7--15.5). Appendix~\ref{app:population} reports how the failure cohort was drawn.

\paragraph{Passage content matters at matched prompt length.}
We replaced each added support segment with screened text from another question, matching position and prompt length to within one token. On the 600 eligible questions, annotated addition exceeded donor addition by 30.5 points [26.4, 34.7]. On all 1,190 failures, literal deletion improved on a fresh baseline by 12.7 points [10.7, 14.6]. Replacing the same candidates with matched donor text improved on that baseline by 9.3 points [7.6, 11.1]. Only literal deletion shortened the prompt. Appendix~\ref{app:population} reports the donor-screening results.

\begin{figure}[htbp]
\centering
\includegraphics[width=\linewidth]{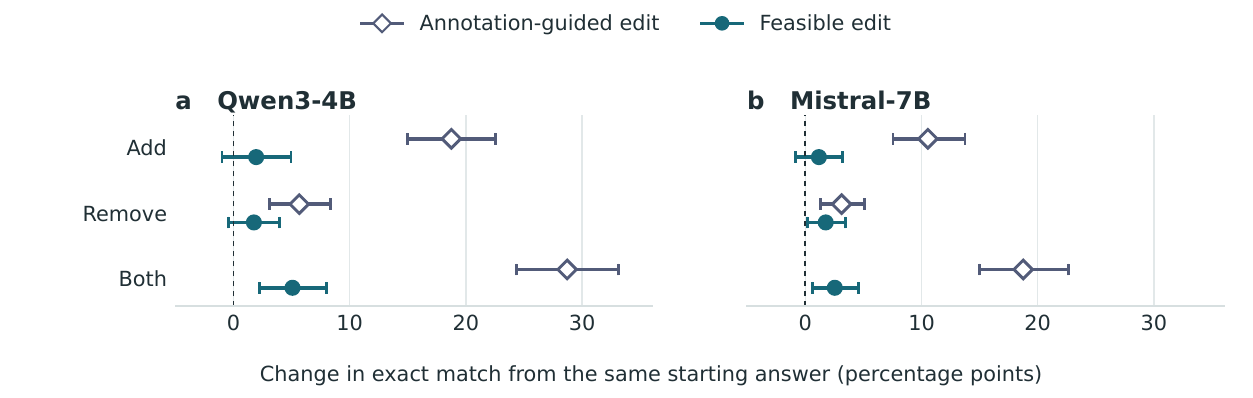}
\caption{Change in exact match relative to the same reader's unedited answer on 512 closed-corpus questions. Each point compares an edited answer and a baseline on the same question. Bars give dataset-stratified, whole-group 95\% intervals. The annotation-guided and feasible edits start with identical retrieval and can select different passages.}
\label{fig:transport}
\end{figure}

\paragraph{Relevance-ranked edits recover smaller gains on both readers.}
On a separate 512-question retrieval test, the joint edit with annotated support raised Qwen exact match by 28.7 points and Mistral exact match by 18.8 points. Choosing additions and deletions by relevance score raised them by 5.1 and 2.5 points, respectively (Figure~\ref{fig:transport}). One bottleneck is the candidate pool. The original top 50 contained 436 of 512 required HotpotQA passages and 394 of 621 required 2Wiki passages. Feasible addition recovered 87 and 50 missing support passages, respectively. Feasible deletion removed two annotated support passages on HotpotQA and one on 2Wiki.

On 128 of these questions we sampled five answers per condition and reader at temperature 0.2. For annotation-guided vectors, exact agreement was 98.75\% within Qwen, 94.92\% within Mistral and 56.41\% across readers. A fixed rule that extracts a leading yes/no answer before scoring raised cross-reader agreement to 60.31\%. With that rule, within-reader agreement was still 98.75\% and 94.61\%, respectively. For Mistral's greedy baseline, the same reference-free extraction credited 68 additional answers among the 98 yes/no questions. These changes show how much of the measured difference can come from answer format. The remaining difference describes the two readers under their respective templates and scoring rules. Appendix~\ref{app:repeats} gives the corresponding feasible-edit comparisons.

\paragraph{Observed edits improve forecasts more readily than action choice.}
We forecasted the exact-match outcome of each feasible edit on 358 initial Qwen failures. The baseline predictor saw the question, the complete original context and the first answer. A second predictor saw these inputs and the three measured annotation-guided outcomes. We excluded a feasible target cell when its rendered prompt equaled any annotation-guided edited prompt. On this comparison the extra measurements lowered mean squared probability error (Brier loss) by .042, .040 and .034 in three separately trained fits. Their respective 95\% intervals were [.034, .051], [.032, .049] and [.027, .042]. These are offline forecasts with access to measured outcomes. A stronger observation representation examined after scoring reduced the incremental gain to .0069 [.0053, .0086].

The count of successful edits and the identity of the edits offer two explanations for this gain. In a follow-up on the same data we fitted the action heads on \emph{training failures}, standardized the observation features on those failures, and chose penalties and model families on \emph{validation failures}. Across five representations, adding the success count reduced Brier loss by .0030--.0071. Adding two orthogonal contrasts identifying which edits succeeded reduced it by a further .0011--.0028. Every individual interval included zero (Appendix~\ref{app:heads}). Training population, feature scaling, and regularization all changed in the refit.

Exact prediction gives a complementary result. Always predicting the most frequent response among training failures gave 52.70\% dataset-macro exact accuracy on held-out failures. The evaluated observation-based predictors achieved 35.48--45.83\% (Appendix~\ref{app:heads}). Several models nevertheless assigned better probabilities to the observed outcomes. For action choice on all 512 questions, always applying the feasible joint edit scored 35.16\%. Direct feasible-action prediction achieved 31.6--34.0\% across seeds. Even selecting the best of the four feasible actions after scoring covered only 205/512 questions, compared with 180/512 for the fixed joint edit. On 307 of the 358 initial failures, every option failed. On those 307 questions, a better selector over the same four actions could not improve exact match. The next experiment gives the system three searches that can retrieve new passages.

\section{Retrieval action selection with reader feedback}
\label{sec:decisions}
\paragraph{The selectors disagree on 45.4\% of decisions.}
With all five original passages restored, targeted deletion and intact diagnosis led to different actions on 1,394 of 3,072 question--seed pairs. Both policies chose from the same menu and returned the chosen action's answer directly. On the 1,024 held-out questions, their exact-match scores were 53.58\% and 53.26\%, a paired difference of $+0.33$ points [$-0.55$, $+1.21$]. After averaging within question over seeds, 58 questions favored targeted feedback, 54 favored intact diagnosis, and 912 tied. The frequent differences in action choice translated into little difference in mean accuracy on this test (Table~\ref{tab:feedback}).

\begin{table}[htbp]
\centering\small
\caption{Feedback for choosing one of five fixed actions. All six selectors use the same menu and return the answer produced by their chosen action. The last column gives the paired target-minus-row difference in official exact match, in percentage points. Intervals are pointwise. Intact diagnosis is the prespecified primary comparator.}
\label{tab:feedback}
\begin{tabular}{@{}lrrl@{}}
\toprule
Selector input beyond the original observation & EM (\%) & F1 (\%) & Target minus row [95\% CI]\\
\midrule
Targeted removal & 53.58 & 60.36 & Reference \\
Diagnosis with intact evidence & 53.26 & 59.91 & $+0.33$ [$-0.55$, $+1.21$] \\
Length-matched random removal & 52.90 & 59.63 & $+0.68$ [$-0.19$, $+1.56$] \\
Unchanged repeat & 52.21 & 58.95 & $+1.37$ [$+0.39$, $+2.34$] \\
Removal metadata only & 52.57 & 59.41 & $+1.01$ [$+0.10$, $+1.93$] \\
Original observation only & 52.60 & 59.52 & $+0.98$ [$+0.05$, $+1.92$] \\
 
\bottomrule
\end{tabular}
\end{table}

\paragraph{Comparison with simpler feedback.}
Targeted feedback improved on an unchanged repeat by $1.37$ points [0.39, 2.34]. Its difference from a length-matched random deletion was $+0.68$ points [$-0.19$, $+1.56$]. In a 257-question panel with stronger initial retrieval, the target-minus-diagnosis difference was $+1.56$ points [$-0.46$, $+3.77$]. Among the 130 questions in this panel whose initial answer was wrong, the difference reached $+3.96$ points [$+0.42$, $+7.87$]. This subset requires reference answers to identify. Its comparisons with random deletion and an unchanged repeat both spanned zero (Appendix~\ref{app:panels}).

\paragraph{The search menu covers 93 more questions on net.}
We counted a question as covered if at least one available action returned an exact match. The four local actions could keep the initial answer, add passages, remove passages, or apply both edits. The five-action menu retained the initial answer and the joint edit, replacing the single edits with three searches. Coverage rose from 566/1,024 (55.27\%) to 659/1,024 (64.36\%), a paired gain of $+9.08$ points [6.97, 11.17]. The five-action menu covered 112 questions missed by all four local actions and missed 19 that the local menu covered (Table~\ref{tab:menus}).

\begin{table}[htbp]
\centering\small
\caption{Which menus contain a correct answer? The first four count columns are mutually exclusive outcomes for a test question. The rescue percentage gives the new-menu successes among old-menu failures. The pooled denominator is 458. The equal-dataset mean is 24.12\%.}
\label{tab:menus}
\begin{tabular}{@{}lrrrrrr@{}}
\toprule
Dataset & Both & Old only & New only & Neither & Questions & Rescue (\%)\\
\midrule
HotpotQA & 299 & 9 & 43 & 161 & 512 & 21.08 \\
2WikiMultiHopQA & 248 & 10 & 69 & 185 & 512 & 27.17 \\
Pooled & 547 & 19 & 112 & 346 & 1,024 & 24.45 \\
 
\bottomrule
\end{tabular}
\end{table}

Of the 458 questions missed by the local menu, the five-action menu recovered 112 (24.45\% pooled or 24.12\% with equal dataset weights). A post-hoc breakdown shows where a better selector could help. The seven distinct actions across both menus covered 678 questions (66.21\%). The five-action menu covered 104 questions missed by the fixed challenge search. On 44, an initially correct answer could have been preserved. On the other 60, some action repaired an initially incorrect answer. Adding the direct diagnosis search to the five-action menu gave a hindsight union of 692/1,024 (67.58\%). Each of these ceilings selects using scored outcomes. It measures opportunity available among the collected answers.

\section{Tracing the deployment gap to answer return}
\label{sec:return}
\begin{figure}[htbp]
\centering
\includegraphics[width=\linewidth]{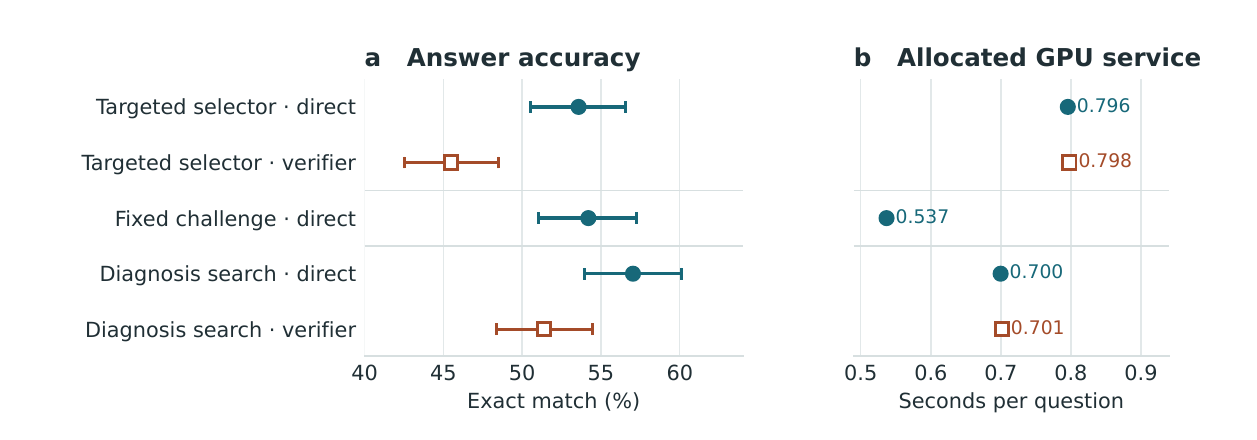}
\caption{Exact match and allocated GPU service for five policies. Filled circles return the chosen action's answer directly. Open squares allow a verifier to override it. The accuracy bars are 95\% group-bootstrap intervals. Appendix~\ref{app:panels}, Table~\ref{tab:deployment}, gives the per-policy reader calls and token counts.}
\label{fig:cost}
\end{figure}

\paragraph{The verifier reduces accuracy on the same selected actions.}
The targeted selector with a verifier scored 45.48\% exact match, against 54.20\% for the validation-selected fixed challenge search. This 8.72-point gap combines two decisions. To locate it, we compared both systems with a third policy after test scoring. This policy used the targeted selector's \emph{same actions} and returned each action's answer directly. It scored 53.58\% (Figures~\ref{fig:first}B and~\ref{fig:cost}).

\begin{table}[htbp]
\centering\small
\caption{Four comparisons on the same 1,024 questions. The first two were specified before test scoring. The last two hold the selected action or the return rule fixed to locate the end-to-end gap. Each estimate is first policy minus second, in percentage points of exact match, with a pointwise 95\% interval.}
\label{tab:contrasts}
\begin{tabular}{@{}llll@{}}
\toprule
What varies & First policy & Second policy & $\Delta$EM [95\% CI]\\
\midrule
Feedback & Targeted, direct & Intact, direct & $+0.33$ [$-0.55$, $+1.21$] \\
End to end & Targeted, verifier & Fixed search, direct & $-8.72$ [$-10.94$, $-6.46$] \\
 
\addlinespace[2pt]
\midrule
Return rule & Targeted, verifier & Targeted, direct & $-8.11$ [$-9.96$, $-6.24$] \\
Policy, direct return & Targeted, direct & Fixed search, direct & $-0.62$ [$-1.98$, $+0.75$] \\
 
\bottomrule
\end{tabular}
\end{table}

On matched actions, the verifier reduced exact match by 8.11 points [6.24, 9.96]. Across the 3,072 question--seed pairs, it turned 323 correct terminal answers into errors and corrected 74 wrong ones, a net loss of 249 correct returns. It returned the initial answer on 36.17\% of question--seed pairs and the temporary probe answer on 24.06\%. With direct return on both paths, the targeted selector scored 0.62 points below fixed challenge search [$-1.98$, $+0.75$ for targeted minus fixed]. The two component differences sum to the end-to-end difference before rounding (Table~\ref{tab:contrasts}). The $+0.33$-point feedback contrast in Section~\ref{sec:decisions} has a different comparator, the intact-diagnosis selector.

\paragraph{The loss also appears on the diagnosis search.}
Executing the intact reader's own proposed query and returning its answer scored 57.03\%. Applying the same verifier to that search lowered accuracy by 5.63 points [3.74, 7.52]. In the direct-return comparison, the targeted selector scored 3.45 points below diagnosis search [$-5.53$, $-1.42$]. Unlike the intact-diagnosis selector in Section~\ref{sec:decisions}, this policy can retrieve with a new query. It was a diagnostic outside the fixed comparator's eligibility pool.

\paragraph{The targeted path also costs more than fixed search.}
On this test, both direct searches had higher mean exact match and used less GPU service than the targeted selector with verification. Fixed challenge search scored 54.20\% at 0.537 GPU seconds per question, and the diagnosis search scored 57.03\% at 0.700. The probe and verifier scored 45.48\% at 0.798 (Figure~\ref{fig:cost}). The fixed challenge search had the highest validation exact match among eligible policies within the probe system's mean validation GPU budget. On test, the targeted selector with verification used 0.261 more allocated GPU seconds per question [0.252, 0.270]. Returning the targeted answer directly still used 0.259 more seconds [0.250, 0.267]. Relative to direct diagnosis search, the difference for the targeted direct path was $+0.096$ seconds [$-0.051$, $+0.173$]. Table~\ref{tab:deployment} gives the validation and test scores for each return rule.

\section{Related work}
RAGChecker evaluates retrieval and generation using fine-grained diagnostics \citep{ru2024ragchecker}. CUE-R perturbs individual evidence items to measure their contribution \citep{jain2026cuer}. Pair-ID asks which edits work together on one question, then compares an annotation-guided measurement with its feasible counterpart from the same initial retrieval. Repeating both readers' answers gives a reference for interpreting disagreement between readers. \citet{mukhopadhyay2026bridge} remove a document and replay the subsequent search trajectory to measure downstream utility. Our temporary deletion elicits one reader response. Restoring the passage before retrieval keeps each action's context fixed across feedback conditions.

Adaptive-RAG chooses a retrieval strategy based on question complexity \citep{jeong2024adaptive}. RePAIR learns to recover after a deficient answer \citep{zhang2026repair}, while counterfactual prompting has also been used to control whether a system should answer at all \citep{rcrag2024}. DynaKRAG learns to select evidence actions as retrieval unfolds \citep{zhou2026dynakrag}. We commit the search queries before feedback and hold each action's terminal answer fixed across feedback conditions. Restoring the passages after the probe isolates the value of observing a response from any direct benefit of leaving the passages edited. Executing the diagnosis's own search changes the set of passages it can access and answers a different operational question.

\section{Limitations}
The experiments use two English multi-hop QA benchmarks. The edit study uses two readers, while the acquisition study uses one reader, one planner, and five actions. The benchmark-derived corpora contain the annotated evidence by construction. Missing documents, changing collections, and longer answers would change the opportunity for retrieval. The recorded service times depend on hardware and batching.

The deletion filters can remove an unannotated bridge passage. No human annotator assessed these passages, so the frequency of this error is unknown. The yes/no extractor tests sensitivity to answer format. The forecasting gains also vary with the representation and the cohort used to fit the prediction heads.

The primary feedback interval includes a small loss and a gain of just over one percentage point. The study specified no equivalence margin. The stronger-retrieval and repeat-feedback panels share questions with the main test, and identifying initially wrong answers requires a reference. The return-rule ablation reused the scored actions and measures the behavior of this fitted verifier.

\section{Conclusion}
Adding three searches to the action menu while removing its two single edits increased coverage by 93 questions on net. The harder problem was identifying those opportunities from the reader's feedback. On the same five actions, targeted deletion and intact diagnosis frequently produced different choices with similar mean exact match. Holding the targeted choices fixed then exposed an 8.11-point loss at answer return. For a retrieval controller with a verifier, retaining the answer produced by each chosen action makes this last failure measurable.

\paragraph{AI assistance.}
OpenAI Codex assisted with implementation, source checking and manuscript drafting and revision. OpenAI ImageGen produced the illustrative artwork. The quantitative marks and intervals use the recorded outcomes and statistical summaries.

{\small\bibliographystyle{plainnat}\bibliography{references}}

\begin{thebibliography}{12}
\providecommand{\natexlab}[1]{#1}
\providecommand{\url}[1]{\texttt{#1}}
\expandafter\ifx\csname urlstyle\endcsname\relax
  \providecommand{\doi}[1]{doi: #1}\else
  \providecommand{\doi}{doi: \begingroup \urlstyle{rm}\Url}\fi

\bibitem[Chen et~al.(2024)Chen, Zhang, Guo, Fan, and Cheng]{rcrag2024}
Lu~Chen, Ruqing Zhang, Jiafeng Guo, Yixing Fan, and Xueqi Cheng.
\newblock Controlling risk of retrieval-augmented generation: A counterfactual
  prompting framework.
\newblock In \emph{Findings of the Association for Computational Linguistics:
  EMNLP 2024}, pages 2380--2393, Miami, Florida, USA, 2024. Association for
  Computational Linguistics.
\newblock \doi{10.18653/v1/2024.findings-emnlp.133}.
\newblock URL \url{https://aclanthology.org/2024.findings-emnlp.133/}.

\bibitem[Ho et~al.(2020)Ho, Nguyen, Sugawara, and Aizawa]{ho2020constructing}
Xanh Ho, Anh-Khoa~Duong Nguyen, Saku Sugawara, and Akiko Aizawa.
\newblock Constructing a multi-hop {QA} dataset for comprehensive evaluation of
  reasoning steps.
\newblock In \emph{Proceedings of the 28th International Conference on
  Computational Linguistics}, pages 6609--6625, 2020.
\newblock \doi{10.18653/v1/2020.coling-main.580}.

\bibitem[Jain and Vedam(2026)]{jain2026cuer}
Siddharth Jain and Venkat~Narayan Vedam.
\newblock {CUE-R}: Beyond the final answer in retrieval-augmented generation.
\newblock \emph{arXiv preprint arXiv:2604.05467}, 2026.
\newblock \doi{10.48550/arXiv.2604.05467}.

\bibitem[Jeong et~al.(2024)Jeong, Baek, Cho, Hwang, and
  Park]{jeong2024adaptive}
Soyeong Jeong, Jinheon Baek, Sukmin Cho, Sung~Ju Hwang, and Jong Park.
\newblock {Adaptive-RAG}: Learning to adapt retrieval-augmented large language
  models through question complexity.
\newblock In \emph{Proceedings of the 2024 Conference of the North American
  Chapter of the Association for Computational Linguistics: Human Language
  Technologies (Volume 1: Long Papers)}, pages 7036--7050, Mexico City, Mexico,
  2024. Association for Computational Linguistics.
\newblock \doi{10.18653/v1/2024.naacl-long.389}.
\newblock URL \url{https://aclanthology.org/2024.naacl-long.389/}.

\bibitem[Jiang et~al.(2023)Jiang, Sablayrolles, Mensch, Bamford, Chaplot,
  de~las Casas, Bressand, Lengyel, Lample, Saulnier, Renard~Lavaud, Lachaux,
  Stock, Le~Scao, Lavril, Wang, Lacroix, and El~Sayed]{jiang2023mistral}
Albert~Q. Jiang, Alexandre Sablayrolles, Arthur Mensch, Chris Bamford,
  Devendra~Singh Chaplot, Diego de~las Casas, Florian Bressand, Gianna Lengyel,
  Guillaume Lample, Lucile Saulnier, L{\'e}lio Renard~Lavaud, Marie-Anne
  Lachaux, Pierre Stock, Teven Le~Scao, Thibaut Lavril, Thomas Wang,
  Timoth{\'e}e Lacroix, and William El~Sayed.
\newblock {Mistral 7B}.
\newblock \emph{arXiv preprint arXiv:2310.06825}, 2023.
\newblock \doi{10.48550/arXiv.2310.06825}.

\bibitem[Mukhopadhyay et~al.(2026)Mukhopadhyay, Ghosh, and
  Chatterjee]{mukhopadhyay2026bridge}
Debayan Mukhopadhyay, Utshab~Kumar Ghosh, and Shubham Chatterjee.
\newblock Bridge evidence: Static retrieval utility does not predict causal
  utility in multi-step agentic search.
\newblock \emph{arXiv preprint arXiv:2607.15253}, 2026.
\newblock \doi{10.48550/arXiv.2607.15253}.

\bibitem[Ru et~al.(2024)Ru, Qiu, Hu, Zhang, Shi, Chang, Cheng, Wang, Sun, Li,
  Zhang, Wang, Jiang, He, Wang, Liu, Zhang, and Zhang]{ru2024ragchecker}
Dongyu Ru, Lin Qiu, Xiangkun Hu, Tianhang Zhang, Peng Shi, Shuaichen Chang,
  Jiayang Cheng, Cunxiang Wang, Shichao Sun, Huanyu Li, Zizhao Zhang, Binjie
  Wang, Jiarong Jiang, Tong He, Zhiguo Wang, Pengfei Liu, Yue Zhang, and Zheng
  Zhang.
\newblock {RAGChecker}: A fine-grained framework for diagnosing
  retrieval-augmented generation.
\newblock In \emph{Advances in Neural Information Processing Systems},
  volume~37, pages 21999--22027, 2024.
\newblock \doi{10.52202/079017-0692}.

\bibitem[Xiao et~al.(2024)Xiao, Liu, Zhang, Muennighoff, Lian, and
  Nie]{xiao2023cpack}
Shitao Xiao, Zheng Liu, Peitian Zhang, Niklas Muennighoff, Defu Lian, and
  Jian-Yun Nie.
\newblock {C-Pack}: Packed resources for general {Chinese} embeddings.
\newblock In \emph{Proceedings of the 47th International ACM SIGIR Conference
  on Research and Development in Information Retrieval}, pages 641--649.
  Association for Computing Machinery, 2024.
\newblock \doi{10.1145/3626772.3657878}.

\bibitem[Yang et~al.(2025)Yang, Li, Yang, Zhang, Hui, Zheng,
  et~al.]{yang2025qwen3}
An~Yang, Anfeng Li, Baosong Yang, Beichen Zhang, Binyuan Hui, Bo~Zheng, et~al.
\newblock {Qwen3} technical report.
\newblock \emph{arXiv preprint arXiv:2505.09388}, 2025.
\newblock \doi{10.48550/arXiv.2505.09388}.

\bibitem[Yang et~al.(2018)Yang, Qi, Zhang, Bengio, Cohen, Salakhutdinov, and
  Manning]{yang2018hotpotqa}
Zhilin Yang, Peng Qi, Saizheng Zhang, Yoshua Bengio, William~W. Cohen, Ruslan
  Salakhutdinov, and Christopher~D. Manning.
\newblock {HotpotQA}: A dataset for diverse, explainable multi-hop question
  answering.
\newblock In \emph{Proceedings of the 2018 Conference on Empirical Methods in
  Natural Language Processing}, pages 2369--2380, 2018.
\newblock \doi{10.18653/v1/D18-1259}.

\bibitem[Zhang et~al.(2026)Zhang, Peng, and Weng]{zhang2026repair}
Gongbo Zhang, Yifan Peng, and Chunhua Weng.
\newblock Improving retrieval-augmented generation without taxonomy-based error
  categorization.
\newblock In \emph{Proceedings of the 64th Annual Meeting of the Association
  for Computational Linguistics (Volume 2: Short Papers)}, pages 155--165, San
  Diego, California, 2026. Association for Computational Linguistics.
\newblock \doi{10.18653/v1/2026.acl-short.14}.
\newblock URL \url{https://aclanthology.org/2026.acl-short.14/}.

\bibitem[Zhou et~al.(2026)Zhou, Wu, Guo, Huang, Zhang, Zhang, Yu, Shi, Lin, and
  Ge]{zhou2026dynakrag}
Chenyu Zhou, Yaqi Wu, Xiaolei Guo, Jiaqi Huang, Xianfa Zhang, Junxu Zhang, Zhuo
  Yu, Zhubo Shi, Jianghao Lin, and Dongdong Ge.
\newblock {DynaKRAG}: A unified framework for learnable evidence control in
  multi-hop retrieval-augmented generation.
\newblock \emph{arXiv preprint arXiv:2607.06507, version 2}, 2026.
\newblock URL \url{https://arxiv.org/html/2607.06507v2}.

\end{thebibliography}
\clearpage
\appendix
\section{Study populations and retrieval protocol}
\label{app:population}
\begin{table}[H]
\centering\small
\caption{Study inventory. Query counts refer to questions. Panel sizes and historical cohorts overlap as indicated. Validation determines policies. Test scores provide evaluation and, when marked, later descriptive comparisons.}
\label{tab:inventory}
\begin{tabularx}{\linewidth}{@{}p{.22\linewidth}p{.19\linewidth}L@{}}
\toprule
Measurement & Questions & Overlap, use and selection\\
\midrule
Historical edits & 1,200 drawn, 1,190 valid & Sampled from 11,105 eligible failures. The 600 with missing support belong to the 1,190. Regeneration excluded ten whose baseline now passed.\\
Historical controls & 640 quota, 180 repeat & 75 quota and 21 repeat questions overlap the valid historical cohort. New matched-donor controls reused the 1,190. These diagnose edit and decoding effects.\\
Retrieved edit pairs & 1,024/256/512 & Training/validation/test. Both readers share the 512 test questions (421 groups). Their results informed the subsequent acquisition design.\\
Reader repeats & 128 of 512 & Five sampled replicates per reader and cell. The 358 Qwen test failures (294 groups) form the separate forecast cohort for this same 512-question draw.\\
Acquisition choices & 1,024/256/1,024 & Reuses development partitions. The test, 1,024 in 871 groups, excludes historical and development groups. Decisions locked before scoring.\\
Hybrid start & 257 of 1,024 & Same questions with different initial retrieval, 223 groups. Model transfers from the main state.\\
Feedback repeats & 128 of 1,024 & Same main-state questions, 115 groups. Three feedback draws conditional on one committed query pair.\\
Post-hoc analyses & Existing draws & Failure-conditioned heads and format checks reuse the edit study. Direct-answer comparisons and menu decompositions reuse the acquisition test.\\
\bottomrule
\end{tabularx}
\end{table}
Of 19,981 benchmark development questions, 11,105 qualified for sampling and 1,200 were selected by hash. Regenerating the baseline left 1,190 failures. In the matched-length controls, a later inference-model check flagged 43 of 4,758 donor blocks as entailing a supplied hypothesis. Seven more exceeded its input limit. All were retained in the reported comparisons.

The four-reader historical replication drew 600 questions from the original 1,200. Each reader's factor effects condition on that reader also failing the baseline. Historical response classes use the archived alias lists and the study's Unicode- and number-normalized exact-match scorer. Closed-corpus runs used the official normalized exact-match and token-F1 scorers, with the released entity aliases and demonyms for 2Wiki. For this reason, the two studies' absolute repair rates should be read within their respective cohorts.

Each benchmark-derived corpus is deduplicated by normalized title and full normalized paragraph, retaining text variants for a title. HotpotQA contributes 507,494 passages and 2Wiki contributes 384,857. BM25 uses $k_1=1.2$ and $b=0.75$. The feasible passage scorer uses 512-token windows with 64 passage-token overlap and the maximum window score. For wider retrieval, BGE-base-en-v1.5 encodes complete paragraphs in windows, and exact inner-product search and BM25 are fused by reciprocal rank before reranking. The same frozen search and prompt construction are used across feedback arms. Empty edits and equal terminal prompts share their generated result. In replay, the system still pays for dependencies that it had to execute before it knew that two prompts coincided.

\section{Reader agreement and answer extraction}
\label{app:repeats}
For a fixed question, let $p_r(v)$ denote the probability of response vector $v$ for reader $r$ under repeated sampling. Within-reader agreement is $\sum_v p_r(v)^2$. Between-reader agreement for $r,s$ is $\sum_v p_r(v)p_s(v)$. We estimate the first from distinct replicate pairs and the second from all cross-reader pairs. Their difference, averaged over the two within-reader agreements, estimates $\frac12\sum_v(p_r(v)-p_s(v))^2$. Five samples per question estimate repeatability conditional on these prompts and temperature. They do not multiply the number of test questions.

\begin{table}[htbp]
\centering\small
\caption{Percent exact vector agreement on 128 paired questions. The fixed extractor removes paired outer quotes from the first nonempty answer line and isolates a leading standalone yes/no followed by punctuation or end of line. It has no access to a reference.}
\label{tab:repeat}
\begin{tabular}{@{}llrrr@{}}
\toprule
Edit family & Scoring & Within Qwen & Within Mistral & Between readers\\
\midrule
Annotation-guided & Official & 98.75 & 94.92 & 56.41 \\
Annotation-guided & Fixed extraction & 98.75 & 94.61 & 60.31 \\
Feasible & Official & 98.12 & 96.25 & 71.06 \\
Feasible & Fixed extraction & 97.81 & 96.25 & 72.62 \\
 
\bottomrule
\end{tabular}
\end{table}
For Mistral's greedy baseline on the 512-question test, extraction raised the count of correct answers from 71 to 139. All 68 changes came from the 98 yes/no questions. For the annotated repeats under official scoring, the within-Qwen, within-Mistral and cross-reader 95\% agreement intervals were [97.2, 100.0], [91.8, 97.7] and [47.8, 64.5] percent. Among the 86 questions on which both readers initially failed, their cross-reader agreement was 67.6\% for annotated edits and 86.5\% for feasible edits. Conditioning and scoring both change what exact agreement measures.

\section{What is predicted by the original observation?}
\label{app:heads}
Let $O$ contain the full question, initial context and original reader response. For a feasible edit $j$, we forecast its success probability $p_j$ and use $(p_j-Y_j)^2$. Each question contributes its average over action cells with rendered input distinct from all annotation-guided edited inputs. On 358 Qwen failures, five feasible-add and nine feasible-delete cells equaled such an input. All three feasible edits were distinct on 344 questions. The original training procedure selected frozen or fine-tuned BGE on validation failures. Its per-action heads used a common regularization grid; they were fitted on eligible training cells from the broader training population. The more informative input adds the observed $V$. A policy that must infer $V$ first faces a different prediction problem.

\begin{table}[htbp]
\centering\small
\caption{Predicting the annotation-guided vector on 358 initial Qwen failures. Exact accuracy and multiclass Brier loss are equal-dataset means. Macro-F1 averages the eight response classes on pooled questions. Empty predicted classes receive zero F1.}
\label{tab:baselines}
\begin{tabular}{@{}lrrr@{}}
\toprule
Predictor & Exact (\%) & Brier loss & Eight-class macro-F1\\
\midrule
All-training majority ($111$) & 4.43 & 1.911 & 0.010 \\
Failure-training majority ($000$) & 52.70 & 0.946 & 0.087 \\
Frozen BGE & 35.48 & 0.818 & 0.145 \\
Qwen+BGE & 45.83 & 0.782 & 0.179 \\
Fine-tuned BGE, seed 1711 & 36.73 & 0.753 & 0.145 \\
Fine-tuned BGE, seed 1729 & 38.36 & 0.732 & 0.142 \\
Fine-tuned BGE, seed 1747 & 44.92 & 0.682 & 0.163 \\
 
\bottomrule
\end{tabular}
\end{table}

The all-training mode is $111$ and performs poorly on failures. The failure-training mode is $000$, occurring on 302/687 training failures and 190/358 test failures. Its pooled success is 53.07\%, while the dataset-macro estimate in Table~\ref{tab:baselines} is 52.70\%. A predictor can assign better probabilities to rarer patterns and improve Brier loss while missing the most frequent exact vector on more questions.

For the failure-conditioned refit, write $K=V_A+V_D+V_{AD}$ and separate count and edit-specific directions as
\[
 u=K/\sqrt3,\qquad c_1=(V_A-V_D)/\sqrt2,\qquad
 c_2=(V_A+V_D-2V_{AD})/\sqrt6 .
\]
In the original coordinates, representing $bK$ with three coefficients incurs squared norm $3b^2$, versus $b^2$ with one count feature. The refit therefore tunes observation and response penalties separately. It also estimates its scaler and action heads using initial training failures and chooses the hyperparameters on initial validation failures, excluding cross-prompt aliases in each action. Validation can select a count-only family. It retained directions for all five representations. Table~\ref{tab:brier} gives the two increments within these matched refits.

\begin{table}[htbp]
\centering\small
\caption{Post-hoc matched refits on existing edit-study data. Entries give Brier improvement and paired 95\% intervals on 358 failures in 294 groups. Lower prediction loss gives a positive gain.}
\label{tab:brier}
\begin{tabular}{@{}lll@{}}
\toprule
Observation representation & Add success count & Add edit directions to count\\
\midrule
Qwen+BGE & $+0.0030$ [$-0.0046$, $+0.0104$] & $+0.0017$ [$-0.0022$, $+0.0058$] \\
Frozen BGE & $+0.0054$ [$-0.0019$, $+0.0127$] & $+0.0011$ [$-0.0024$, $+0.0047$] \\
Fine-tuned BGE, 1711 & $+0.0041$ [$-0.0044$, $+0.0126$] & $+0.0028$ [$-0.0008$, $+0.0066$] \\
Fine-tuned BGE, 1729 & $+0.0071$ [$-0.0016$, $+0.0157$] & $+0.0015$ [$-0.0021$, $+0.0051$] \\
Fine-tuned BGE, 1747 & $+0.0057$ [$-0.0020$, $+0.0135$] & $+0.0024$ [$-0.0012$, $+0.0062$] \\
 
\bottomrule
\end{tabular}
\end{table}

\section{Feedback controls, test strata and cost}
\label{app:panels}
All feedback selectors used the same question, initial passages, initial answer, and committed gap and challenge queries. Targeted and random deletion used a temporary edit. The random control selected a length-matched alternative, breaking ties by a fixed hash. The unchanged repeat sampled a response to the intact passages. The operation-only control exposed the target metadata while withholding the response. All response-producing feedback conditions sampled at temperature 0.2. The terminal answer was greedy for a given rendered context. The ancillary files include the reader and planner prompts and the retrieval parameters.

Before choosing, a selector can see the question, the original passages and scores, the initial answer, the committed query strings, its own feedback, and predefined features of the response and operation. References, support annotations, probe correctness and unselected search results are withheld. All six information conditions share one table of action outcomes. Identical terminal prompts reuse a single greedy response.

For each training fraction and seed, validation first selects a representation of the original observation from a frozen Qwen+BGE encoding and trained BGE checkpoints. The criterion is dataset-macro exact match for five-action choice, with ties broken by Brier loss and then cost. Given that representation, each feedback condition fits five action-success heads, selecting logistic or multilayer perceptron models on validation. The labels for a given question and action are shared across conditions. Validation accuracy and cost also select the fixed comparator. Eligible policies include the two-answer selector, verified union, verified diagnosis and fixed actions with each return rule. Direct diagnosis and direct union are prespecified diagnostics outside this pool. The shared verifier is fitted on training candidates.

\begin{table}[htbp]
\centering\small
\caption{Targeted deletion minus diagnosis on intact passages. Official exact-match differences are in percentage points. Subsets overlap. Every interval is pointwise. All values except the half-training and per-seed rows use the full training partition and all three trained policies.}
\label{tab:panels}
\begin{tabular}{@{}lrrl@{}}
\toprule
Subset & Questions & Groups & Difference [95\% CI]\\
\midrule
Main state & 1,024 & 871 & $+0.33$ [$-0.55$, $+1.21$] \\
Half of training questions & 1,024 & 871 & $+0.42$ [$-0.42$, $+1.27$] \\
HotpotQA & 512 & 464 & $+0.26$ [$-1.11$, $+1.65$] \\
2WikiMultiHopQA & 512 & 407 & $+0.39$ [$-0.65$, $+1.44$] \\
Initially correct & 445 & 419 & $+0.48$ [$-0.58$, $+1.57$] \\
Initially wrong & 579 & 492 & $+0.24$ [$-1.12$, $+1.64$] \\
All support in top 50 & 521 & 491 & $+0.15$ [$-0.89$, $+1.21$] \\
Incomplete support in top 50 & 503 & 394 & $+0.55$ [$-1.24$, $+2.34$] \\
Yes/no answer & 219 & 219 & $-0.95$ [$-3.29$, $+1.23$] \\
Other answer & 805 & 677 & $+0.46$ [$-0.46$, $+1.39$] \\
Hybrid initial state & 257 & 223 & $+1.56$ [$-0.46$, $+3.77$] \\
Hybrid, initially wrong & 130 & 112 & $+3.96$ [$+0.42$, $+7.87$] \\
Main, training seed 1811 & 1,024 & 871 & $-0.29$ [$-1.76$, $+1.18$] \\
Main, training seed 1829 & 1,024 & 871 & $+1.17$ [$-0.29$, $+2.63$] \\
Main, training seed 1847 & 1,024 & 871 & $+0.10$ [$-1.08$, $+1.27$] \\
 
\bottomrule
\end{tabular}
\end{table}

On the three-draw feedback panel, the targeted selector chose the same action in 96.70\% of within-question draw pairs [94.71, 98.35]. The diagnosis selector agreed in 93.92\% [91.84, 95.88]. The labels needed to identify the initially incorrect subsets in Table~\ref{tab:panels} were used for analysis only. The original single-edit diagnostics were collected on the main initial state only, so the old four-action coverage has no stronger-state estimate.

\paragraph{Costs at different levels.}
The main policy estimates allocate measured GPU batch service to a question on the selected path. They charge retrieval and reranking, observation encoding, reader calls, and verification when incurred. The full timing suite is a separate measurement on 16 validation questions, running twelve slots per question. It averaged 10.54 wall-clock seconds per question in the single-query profile and 2.95 in the throughput profile (shared block time allocated per question). Neither quantity is the latency of an individual policy.

The ancillary data tables report unrounded intervals, recorded costs and question-level outcomes. In its policy tables, \pname{INFO\_TARGET} and \pname{INFO\_GAP} denote the targeted and intact-diagnosis information selectors. \pname{PROBE\_REUSE} applies the verifier to the targeted path. \pname{FIXED\_SELECTED} is the fixed challenge search ($A_4$). \pname{FREE\_GAP\_RAW} and \pname{FREE\_GAP\_VERIFY} execute the diagnosis's own query and use direct and verifier return, respectively.

\begin{table}[H]
\centering\small
\caption{End-to-end policies on the 1,024-question test. The validation scores were available at policy selection. Costs are per question on the selected logical path, including the baseline and any planning, feedback and terminal calls. GPU seconds are allocated batch service. Input and output columns count reader tokens.}
\label{tab:deployment}
\begin{tabular}{@{}lrrrrrr@{}}
\toprule
Policy and return rule & Val.\ EM & Test EM & GPU s & Calls & Input & Output\\
 & (\%) & (\%) & & & tokens & tokens\\
\midrule
Targeted selector, direct & 53.91 & 53.58 & 0.796 & 3.90 & 2,558 & 55.2 \\
Targeted selector, verifier & 38.28 & 45.48 & 0.798 & 3.90 & 2,558 & 55.2 \\
Fixed challenge, direct & 52.34 & 54.20 & 0.537 & 3.00 & 2,113 & 51.1 \\
Diagnosis search, direct & 57.03 & 57.03 & 0.700 & 3.00 & 2,132 & 75.7 \\
Diagnosis search, verifier & 45.83 & 51.40 & 0.701 & 3.00 & 2,132 & 75.7 \\
 
\bottomrule
\end{tabular}
\end{table}

\FloatBarrier
\end{document}